# Geophysical tomography in engineering geology: an overview

## Domenico Patella

*Department of Physical Sciences, University Federico II, Naples, Italy (E-mail: patella@na.infn.it)*

## ABSTRACT

An overview of the tomographic interpretation method in engineering geophysics is presented, considering the two approaches of the deterministic tomography inversion, developed for rock elasticity analysis, and the probability tomography imaging developed in the domain of potential fields methods. The theoretical basis of both approaches is shortly outlined before showing a laboratory and a field application.

## INTRODUCTION

Geophysical prospecting is widely applied to help resolve many problems in civil and environmental engineering. The probability of a successful application at a given site rapidly increases if different methods are used, basing the selection on the principles of complementarity and coherency of information. Such a strategy is mainly advisable in delicate environments, where absolutely non-intrusive geophysics is the only possibility for target identification, prior to direct exploration works [1].

The interpretation of geophysical datasets to elaborate accurate images of the investigated structures has always been a difficult task, due to the mathematical difficulties and heavy calculations involved in modeling approaches. Favourably, the newest hardware and software generation has raised to such a high level of sophistication to allow, at last, a routine application of the complex imaging tools so far developed.

In the following sections an outline of the tomography imaging in engineering geophysics is reported, focussed on the two approaches of the deterministic tomography (DT), developed in rock elasticity, and the probabilistic tomography (PT), proposed for potential field methods.

## DETERMINISTIC TOMOGRAPHY INVERSION

The DT inversion in seismic prospecting was developed to derive a detailed geometrical model of the elastic wave velocity pattern in a medium, starting from the measured wave travel times from sources to receivers [2] (fig.1).

The relationship relying the wave travel time $t_i$ to the unknown slowness function $s(x,y,z)$ for a ray along the $i$-th path $l_i$ of a set of $M$ paths, is given by Fermat integral

$$t_i = \int_{l_i} s(x,y,z)dl , \quad (i=1,2,..,M). \tag{1}$$

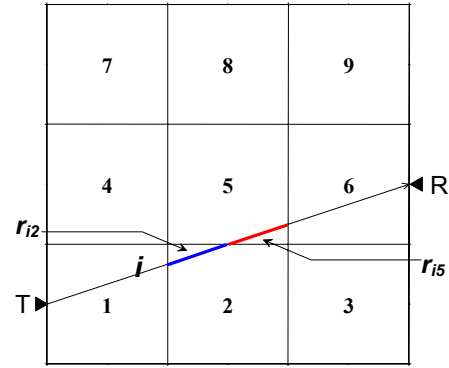

**Figure 1** A transmitter-receiver array for the application of the deterministic tomography inversion.

By defining the delay time as the difference between the measured travel time and the travel time in an a priori assigned reference uniform medium and subdividing the medium into $N$ elementary cells, with the application of a perturbation technique it is possible to deduce from eq.1 a system of linear equations in matrix form as

$$\delta \mathbf{t} = \mathbf{R} \cdot \delta \mathbf{s}, \tag{2}$$

where $\delta \mathbf{t}$ is a column matrix, whose element $\delta t_i$ ($i=1,..,M$) is the delay time along the $i$-th path, $\mathbf{R}$ is a rectangular sparse matrix, whose element $r_{ij}$ ($i=1,..,M; j=1,..,N$) is the path length of the $i$-th ray in the $j$-th cell, and $\delta \mathbf{s}$ is a line matrix, whose element $\delta s_j$ ($j=1,..,N$) is the slowness departure from the reference model in the $j$-th elementary cell.

If the space of the experimental data has dimensions much greater than those of the space of the unknowns (*M»N*), the system (2) becomes overdetermined. A least-squares procedure can thus be applied, by minimizing the Euclidean norm $\|\mathbf{R}\cdot\delta\mathbf{s}-\delta\mathbf{t}\|$. The solution for the vector $\delta\mathbf{s}$ is then given as

$$\delta\mathbf{s}=(\mathbf{R}^T\mathbf{R})^{-1}\mathbf{R}^T\delta\mathbf{t}=\mathbf{A}^{-1}\mathbf{R}^T\delta\mathbf{t}. \tag{3}$$

In solving eq.3 one may encounter serious difficulties, essentially related to the presence of small values in the matrix to invert. Thus, it is preferable to constrain matrix **A** by introducing a damping factor $\beta$ and use as solution for $\delta\mathbf{s}$ the equation

$$\delta\mathbf{s}=(\mathbf{A}-\beta\mathbf{I})^{-1}\mathbf{R}^T\delta\mathbf{t}, \tag{4}$$

where **I** is the identity matrix.

To increase the resolution power of the DT method an iterative procedure can be applied, consisting in using the slowness model drawn from an inversion as the reference model for a new inversion. Iterations will be stopped as soon as the mean departure between the slowness values got from the *k*-th and (*k*-1)-th iterations is in modulus not greater than a pre-fixed discrepancy factor $\Delta$, *i.e.*

$$\left|\frac{1}{N}\sum_{j=1}^{N}\left(s_j^k - s_j^{k-1}\right)\right| \leq \Delta. \tag{5}$$

### PROBABILISTIC TOMOGRAPHY IMAGING

The purpose of the PT procedure is to retrieve an image of the spatial distribution of the occurrence probabilities of the sources of the observed anomalies [3,4].

Consider a reference coordinate system with the (*x*,*y*)-plane at sea level and the *z*-axis positive upwards, and a survey surface *S* with uneven topography (see fig.2). Let $A(\mathbf{r})$ be the anomaly value at a station located at $\mathbf{r}\equiv(x,y,z)$, with $\mathbf{r}\in S$, and assume that it can be discretized as a sum of partial effects due to *Q* elementary sources, viz.

$$A(\mathbf{r}) = \sum_{q=1}^{Q} a_q s(\mathbf{r}-\mathbf{r}_q), \tag{6}$$

The *q*-th elementary source, located at $\mathbf{r}_q\equiv(x_q,y_q,z_q)$, is given a strength $a_q$ and its effect at the station at $\mathbf{r}\equiv(x,y,z)$ is analytically described by the kernel $s(\mathbf{r}-\mathbf{r}_q)$.

The information power $\Lambda$ over *S* associated with $A(\mathbf{r})$ is defined as

$$\Lambda = \iint_S [A(\mathbf{r})]^2 dS, \tag{7}$$

which using eq.6 can be made explicit in the form

$$\Lambda = \sum_{q=1}^{Q} a_q \iint_S A(\mathbf{r}) \cdot s(\mathbf{r}-\mathbf{r}_q) dS. \tag{8}$$

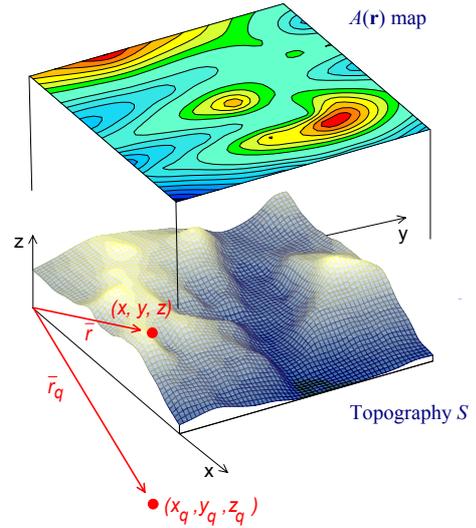

**Figure 2** The conceptual assumptions for the application of the probabilistic tomography imaging.

Consider a generic *q*-th addendum in eq.8 and apply Schwarz's inequality, thus obtaining

$$\left[\iint_S A(\mathbf{r}) \cdot s(\mathbf{r}-\mathbf{r}_q) dS\right]^2 \leq$$
$$\leq \iint_S A^2(\mathbf{r}) dS \cdot \iint_S s^2(\mathbf{r}-\mathbf{r}_q) dS. \tag{9}$$

Using the inequality (9), a *source element occurrence probability* (SEOP) function is at last defined as

$$\eta(\mathbf{r}_q) = C_q \iint_S A(\mathbf{r}) s(\mathbf{r}-\mathbf{r}_q) dS \tag{10}$$

where

$$C_q = \left[\iint_S A^2(\mathbf{r}) dS \iint_S s^2(\mathbf{r}-\mathbf{r}_q) dS\right]^{-1/2}. \tag{11}$$

The SEOP function meets the condition $-1 \leq \eta(\mathbf{r}_q) \leq +1$, and is interpreted as a measure of the probability, which a source element with strength $a_q$ placed at $\mathbf{r}_q$ obtains as responsible of the whole observed anomaly field $A(\mathbf{r})$.

The PT procedure for a dataset collected on a non-flat topography $S$ consists in a scanning procedure based on the knowledge of the $s(\mathbf{r}-\mathbf{r}_q)$ function that is called *space domain scanner*. It is a function depending on the method used for sensing the earth and is generally well defined, since it represents the physical behaviour of the field due to a source element (*e.g.* resistivity element [5], electrical current filament [6], gravitational mass [7], *etc.*).

In practice, as the true source distribution responsible of an observed anomaly field $A(\mathbf{r})$ is unknown, a positive source element of unitary strength can be used to scan the exploration volume (the tomospace) to search where the sources are most probably located. For any tern $x_q, y_q, z_q$ in the tomospace, the integral (10) gives the probability that a positive ($\eta>0$) or negative source ($\eta<0$) located in that point is responsible for the $A(\mathbf{r})$ field detected on surface. By scanning the tomospace along a sequence of slices, a 3D image reconstruction of the sources distribution can be finally drawn in a probabilistic sense.

## APPLICATIONS

### A DT laboratory experiment

Fig.3 depicts a schematic planar section of a (32×32×6) $cm^3$ composite block, consisting of a roughly (11×16×5) $cm^3$ piece of marble ($v_p$=4.6 $km/s$) buried within a chalky matrix ($v_p$=1.6 $km/s$). The measurements were done using a Panametrics 5058PR ultrasonic impulse generator, two Panametrics X1021 *p*-wave transducers with a resonance central frequency at 50 *Hz*, and a Tektronix TDS430A digital oscilloscope.

The block was ideally subdivided in 64 equal cells of volume (4×4×6) $cm^3$. The measuring procedure consisted each time in fixing the pair of transmitter and receiver at the centre of the vertical facelets of area (4×6) $cm^2$, lying along the opposite faces of the block, and moving one or both transducers at the constant step of 4 *cm*, as shown in fig.3. In order to improve resolution, the transmitter and receiver roles were interchanged, thus obtaining a total of *M*=256 ray paths

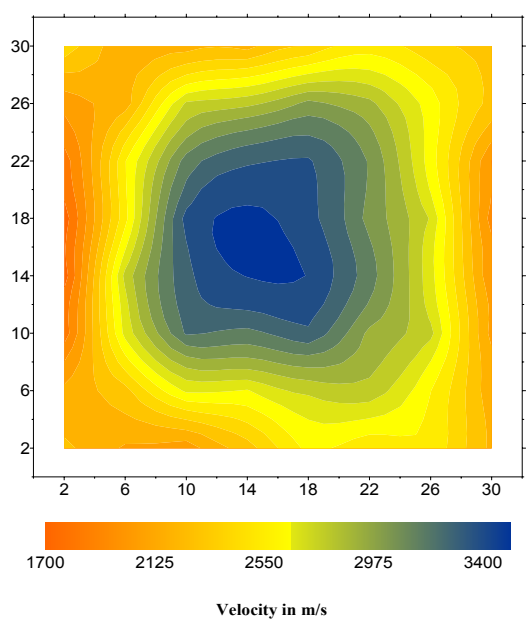

**Figure 4**    Example of a laboratory application of the DT inversion. Results of the iterated DT algorithm.

The results obtained from the application of the DT algorithm are shown in the 2D map of fig.4. A damping factor $\beta$=0.1 and a discrepancy factor $\Delta$=1 were assumed. A clear conformity appears between the irregular form of the piece of marble and its reconstructed physical image. This picture, jointly with the reasonable estimate of the velocity profile, demonstrates how high is the resolution this DT technique can reach, at least in laboratory.

### A PT field experiment

Geoelectrics is often utilised to get resistivity information near and over waste disposal sites. It helps mapping both the vertical and horizontal distribution of contamination caused by increase of solutes in groundwater relative to background levels, which is reflected in some increase of the electrical conductivity of the water bearing rock [8].

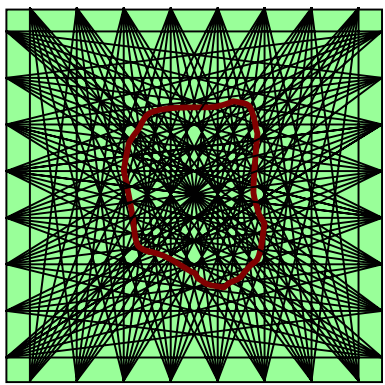

**Figure 3**    A laboratory application of the DT inversion. Sample trace and transducers layout.

Dipole-dipole profiling is the most adopted technique, as it provides high vertical and lateral sensitivity. The so called pseudo-section representation allows a preliminary inspection to be made in terms of apparent resistivity.

Figs.5 and 6 refer to a field case performed on a waste disposal site, consisting of a waterproofed basin dug out in a sandy-clayey layer down to 17 *m* of depth b.g.l.. The basin was wholly filled with wastes and leaks of pollutant were suspected across tears in the impermeable sheets of 5 *mm* of thickness.

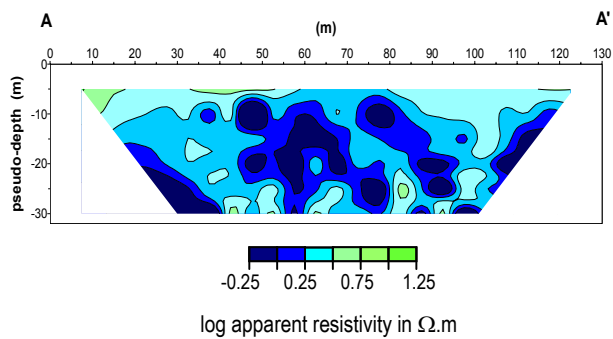

**Figure 5**  Example of a field application of the PT imaging. A dipole-dipole pseudo-section.

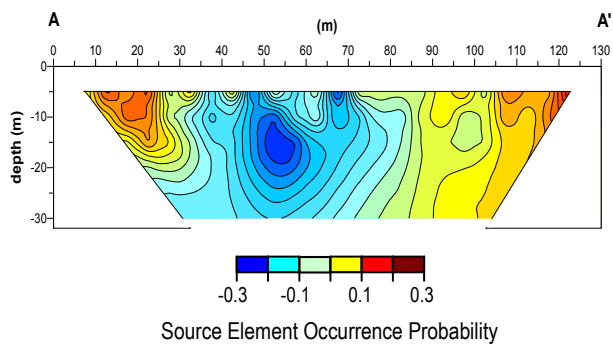

**Figure 6**  Example of a field application of the PT. Results from the PT algorithm.

The pseudo-section in fig.5 shows apparent resistivity variations in the range 0.8 - 10 $\Omega m$. A general uniformity characterizes the lateral portions of the profile, where the apparent resistivity appear to increase with depth. The central part of the section shows, instead, very low values of about 1 $\Omega m$ along the whole depth scale with presence of small nuclei enclosing even lower values.

Fig.6 shows the results of the PT algorithm applied to the pseudo-section of fig.5. The most remarkable feature is the presence of the lowest negative values of the SEOP function at the left-hand border of the central part of the section. In particular, the largest negative SEOP nucleus, located between 40 and 70 *m* along the horizontal profile, appears to propagate well beyond the impermeable sheet. The conclusion is thus that the pollutant solutes may have overstepped the barrier, though limited to a short distance thanks to the low permeability nature of the sandy-clayey hosting deposit.

## CONCLUSION

An overview of the tomographic interpretation method in engineering geophysics has been given. Two approaches have been outlined, namely the deterministic tomography (DT) inversion, developed for acoustic velocity analysis in rocks, and the probability tomography (PT) imaging developed for electrical resistivity analysis in the subsoil. The theoretical principles of both approaches have been briefly reported and two case-studies have been analysed in order to demonstrate the highest resolution power that tomography provides in geophysics.